\documentclass[%
 reprint,
 amsmath,amssymb,
 aps,
]{revtex4-1}

\usepackage{graphicx}
\usepackage{dcolumn}
\usepackage{bm}
\usepackage{subfigure}

\begin{document}

\preprint{APS/123-QED}

\title{Induced p-wave superconductivity without spin-orbit interactions}

\author{Fernanda Deus}
\author{Mucio A. Continentino}
\affiliation{Centro Brasileiro de Pesquisas F\'{i}sicas \\
Rua Dr. Xavier Sigaud, 150, Urca, Rio de Janeiro, RJ 22290-180, Brazil}

\author{Heron Caldas}
\affiliation{Departamento de Ci\^{e}ncias Naturais, Universidade Federal de S\~{a}o Jo\~{a}o Del Rei, S\~{a}o Jo\~{a}o Del Rei, MG, Brazil\\}

\date{\today}

\begin{abstract}
The study of Majorana fermions  is of great importance for the implementation of a quantum computer. These modes are topologically protected and very stable. It is now well known that a $p$-wave superconducting wire can sustain, in its topological non-trivial phase, Majorana quasi-particles at its ends. Since this type of superconductor is not found in nature, many methods have been devised to implement it. Most of them rely on the spin-orbit interaction. 
In this paper we study the  superconducting properties of a two-band system in the presence of antisymmetric hybridization. We consider inter-band attractive interactions and also an attractive interaction in one of the bands. We show that superconducting fluctuations with $p$-wave character are induced in the non-interacting band due to the combined effects of inter-band coupling and hybridization. In the case of a wire, this type of induced superconductivity gives rise to four Majorana modes at its ends. The long range correlation between the different charge states of these modes   offers  new possibilities for the implementation of protected $q$-bits. 


\end{abstract}

\pacs{Valid PACS appear here}

\maketitle

\section{\label{intro}Introduction}

In metallic multi-band systems, electrons arising from different atomic orbitals coexist at a common Fermi surface. Electronic states, either in the same or different sites may overlap and mix through the crystalline potential giving rise to hybrid bands.  Superconductivity is strongly affected by hybridization~\cite{petrovic, ramos, dani, daniel, bauer2, scheila, bauer}. In general this has a detrimental effect and can even destroy it at a superconducting quantum critical point (SQCP) \cite{daniel}. This has been verified experimentally~\cite{bauer2, scheila, bauer} in multi-band superconductors which are driven to the normal state by external pressure which increases the overlap of the wave functions and consequently their mixing.  Theoretically~\cite{igor, mucio} this has been verified for the case of a symmetric, $k$-independent hybridization. However, in many cases, as will be discussed in this paper, hybridization can be antisymmetric~\cite{takagi, julien, aline} and we should consider its $k$-dependence. Here, we study the influence of odd-parity mixing on  the superconducting properties of a multi-band superconductor. We show it has non-trivial effects in the superconducting properties. We demonstrate that antisymmetric hybridization can enhance superconductivity~\cite{mucio2} in multi-band systems. We also show that it can promote a crossover from a weak coupling, Bardeen, Cooper and Schrieffer (BCS)~\cite{aproxbcs} type of superconductivity to a strong coupling one associated with a Bose-Einstein condensation (BEC) of pairs~\cite{igor, mucio}.

One of the most remarkable properties of a one dimensional $p$-wave superconductor was discovered by Kitaev \cite{kitaev}; he has shown that Majorana fermions can exist at the ends of this system.  The idea of Majorana fermions was introduced by Ettore Majorana in 1937 \cite{majorana}. After more then 70 years it was proposed, there is still no definitive way to detect it \cite{wilczek, stern, franz, service, hughes}. This particle has the eccentric property of being its own antiparticle \cite{ralicea}. In condensed matter systems Majorana fermions are emergent quasi-particles, topologically protected and that satisfy a criterion of robustness for use in quantum computers. They are promising candidates to acts as $q$-bits \cite{ladd}. In the high-energy context, there is a current idea that neutrinos may be Majorana fermions~\cite{avignone}. The investigation of Majorana fermions is also important in the context of fundamental physics. 

It becomes clear that obtaining Majorana fermions is of great importance. Motived by this, we propose here a new mechanism to produce a $p$-wave one-dimensional superconductor, without the necessity of spin-orbit interactions and an external magnetic field. The two-band Hamiltonian we study in this paper represents an effective model to describe a non-interacting wire deposited on top of a bulk superconductor. Assuming that the hybridization between the electronic states in the wire and on the bulk is  antisymmetric, we show that the superconductivity induced in the wire has a $p$-wave character. However, since we are dealing with spinful fermions, the pairing we obtain corresponds to the $m_{l}=0$ of the $l=1$ $p$-wave state. This type of pairing does not break time reversal symmetry and we find four Majorana modes in the chain, two at each end. Although two Majorana can combine to form a standard fermion,  these composite Majoranas exhibit highly non-trivial properties~\cite{chines} and their charge states have long range correlations along the wire. 

%

\section{\label{origem}Origin of  Odd-Parity Hybridization}

In this section, we discuss the origin of an antisymmetric hybridization following the work in Ref.~\cite{drzazga}. We assume, as these authors, that the Wannier functions~\cite{wannier, wannier2} of the $s$, $p$, $d$ or $f$ electrons of the solid have the same parities as the corresponding atomic functions.

We consider that hybridization is caused by a periodic lattice potential,  $v(\vec{r})$. This lattice has inversion symmetry, i.e., $v(-\vec{r}) = v(\vec{r})$. The matrix elements of hybridization are written as,
\begin{eqnarray}
V_{l l^{\prime}} (\vec{r}_{1} - \vec{r}_{2}) = \int_{-\infty}^{\infty} d\vec{r}^{\;\prime}  \psi_{l}^{*}(\vec{r}^{\; \prime} - \vec{r}_{1}) v(\vec{r}^{\; \prime}) \psi_{ l^{ \prime}}(\vec{r}^{\; \prime} - \vec{r}_{2}), \nonumber
\end{eqnarray}
where $\psi_{l ( l^{ \prime})}$ is the wave function of $l$$( l^{ \prime})$ orbital. We can take $\vec{r}_{1} = 0$ and $\vec{r}_{2} = - \vec{r}$. So,
\begin{eqnarray}
\label{hibri}
V_{l l^{\prime}} (\vec{r}) = \int_{-\infty}^{\infty} d\vec{r}^{\;\prime}  \psi_{l}^{*}(\vec{r}^{\; \prime}) v(\vec{r}^{\; \prime}) \psi_{ l^{ \prime}}(\vec{r}^{\; \prime} + \vec{r}),
\end{eqnarray}
where we can write $\psi$ in spherical coordinates~\cite{griffiths} as,
\begin{eqnarray}
\label{eqpsi}
\psi_{l} (\vec{r}) = \psi_{lm}(r, \theta, \phi) = R(r) Y_{l}^{m}(\theta, \phi),
\end{eqnarray}
with $R(r)$ the radial solution of Laplace's equation and $Y_{l}^{m}(\theta, \phi)$  the angular solution, known as spherical harmonics. The indexes $l$ and $m$ are quantum numbers, such that $l > 0$ and $m = -l,...,0,...,+l$.

We want to investigate the parity of Eq.~\ref{hibri}. This is possible by doing an inversion of coordinates: $\vec{r} \rightarrow - \vec{r}$. Doing this in equation (\ref{eqpsi}), we notice that the parity of the wave function depends on the parity of the $Y_{l}^{m} $, the spherical harmonics, which depends on $l$ by the following expression,
\begin{eqnarray}
Y_{l}^{m} (\pi - \theta, \pi + \phi) = (-1)^{l} Y_{l}^{m} (\theta, \phi).
\end{eqnarray}
So,
\begin{eqnarray}
\label{eqneg}
&&\psi_{l} (-\vec{r})  = (-1)^{l} \psi_{l} (\vec{r}),
\end{eqnarray}
such that, $\psi_{l}^{*} (-\vec{r}) = (-1)^{-l} \psi_{l}^{*} (\vec{r})$.

Now we are able to determine the parity of Eq.~\ref{hibri}. Performing an inversion of coordinates $\vec{r} \rightarrow - \vec{r}$, we can write,
\begin{eqnarray}
V_{l l^{\prime}}(-\vec{r}) = \int_{-\infty}^{\infty} d\vec{r}^{\;\prime}  \psi_{l}^{*}(\vec{r}^{\; \prime}) v(\vec{r}^{\; \prime}) \psi_{l^{\prime}}(\vec{r}^{\; \prime} - \vec{r}), \nonumber
\end{eqnarray}
doing a  change of variable $\vec{r}^{\;\prime} = - \vec{r}^{\;\prime \prime}$, we get
\begin{eqnarray}
&&V_{l l^{\prime}}(-\vec{r}) = \int_{-\infty}^{\infty} d\vec{r}^{\; \prime \prime}  \psi_{l}^{*}(- \vec{r}^{\;\prime \prime}) v(\vec{r}^{\;\prime \prime}) \psi_{l^{\prime}}\left(- \left[\vec{r}^{\;\prime \prime} + \vec{r}\;\right] \right). \nonumber
\end{eqnarray}
Finally, using Eq.~\ref{eqneg} and doing some simple manipulations, we get,\begin{eqnarray}
&&V_{l l^{\prime}}(-\vec{r}) = (-1)^{l^{\prime} - l} V_{l l^{\prime}} (\vec{r}).
\end{eqnarray}
We can conclude that the parity of the hybridization depends on the difference of the angular momenta of the electronic orbitals, $l^{\prime} - l$. Since,  $l$ is always positive, $V_{l l^{\prime}}(\vec{r})$ has {\it{even parity}} if $l^{\prime} - l$ is an even number and {\it{odd parity}} if $l^{\prime} - l$ is an odd number. Then, every time we mix orbitals in neighboring sites  with angular momentum $l$ and $l + 1$, we need to consider odd-parity hybridization. The antisymmetric relation in real space is $V(- \vec{r}) = - V(\vec{r})$ and in momentum space it is given by, $V (-\vec{k}) = - V (\vec{k})$. In a one-dimensional lattice, for example, an  antisymmetric hybridization is given by, $V (k) = V_k \propto i \sin k a$ with $a$ the lattice spacing.
An additional important constraint is that of time reversal symmetry which implies for the Hamiltonian studied here that the antisymmetric hybridization $V_k$ is a purely imaginary quantity.

Notice that the case of antisymmetric $V(\vec{r})$ is of great relevance for condensed matter physics as it includes the $s$-$p$, hybridization, $d$-$p$  mixing, relevant for the copper oxides,  and $d$-$f$ mixing that encompasses many rare-earth systems, the actinides and their compounds. Antisymmetric mixing is also an essential ingredient to give rise to topological insulating phases in multi-band systems~\cite{spchain}.

\section{\label{modelo}Model}

As we mentioned earlier, we focus our attention on a two-band system, with an odd-parity hybridization between these bands, an attractive (inter-band) interaction between the electrons in different bands, and also  an attractive (intra-band) interaction between electrons {\it in only one of} the bands. The Hamiltonian of this problem can be written as,
\begin{eqnarray}
\label{hamiltonian}
&&H = \sum_{k, \sigma} \left( \epsilon_{k}^{a} a_{k \sigma}^{\dag} a_{k \sigma} + \epsilon_{k}^{b} b_{k \sigma}^{\dag} b_{k \sigma} \right)   \nonumber\\
&&\;\;\;\;- \; \sum_{k \sigma} \left(\Delta_{ab} a_{k \sigma}^{\dag} b_{-k -\sigma}^{\dag} + \Delta_{ab}^{*} b_{-k -\sigma} a_{k \sigma}\right)  \nonumber\\
 &&\;\;\;\;- \;  \sum_{k \sigma}  \left(\Delta_{bb} b_{k \sigma}^{\dag} b_{-k -\sigma}^{\dag}  + \Delta_{bb}^{*} b_{-k -\sigma} b_{k \sigma}\right)   \nonumber\\
&& \;\;\;\;+  \sum_{k \sigma} \left( V_{k} a_{k \sigma}^{\dag} b_{k \sigma}  + V_{k}^{*} b_{k \sigma}^{\dag} a_{k \sigma}\right),
\end{eqnarray}
where $\sigma$ is the spin index that could be ``up'' ($\uparrow $) or ``down'' ($\downarrow $), $\epsilon_{k}^{a,b}$ are the energies of the electrons in the $a$ and $b$  bands. In an obvious notation, $a_{k \sigma}^{(\dag )} $ and $b_{k \sigma}^{(\dag )} $ annihilate (create) electrons in these bands respectively. The attractive many-body term has been decoupled using the BCS approximation~\cite{aproxbcs}. The odd parity hybridization is such that, $V(-\vec{r}) = - V(\vec{r})$ in real space, or in  $k$-space $V_{-k} = - V_{k}$ and  mixes states with the same spin. 

The order parameters that characterize the superconducting phase are, the {\it inter-band} superconducting order parameter given by,
\begin{eqnarray}
\label{inter}
\Delta_{ab}  \equiv  g_{ab} \sum_{k \sigma}  \langle a_{k \sigma} b_{-k -\sigma} \rangle
\end{eqnarray}
 and the {\it intra-band} one,
\begin{eqnarray}
\label{parintra}
\Delta_{bb} \equiv g_{bb}  \sum_{k \sigma} \langle b_{k \sigma} b_{-k -\sigma} \rangle.
\end{eqnarray}
where $g_{ab}$ and $g_{bb}$ are the attractive interactions.
Although there is no attractive interaction in the $a$-band, we will investigate the existence of induced superconductivity  in this band. For this purpose we  define the $k$-dependent anomalous correlation function in the $a$-band:
\begin{eqnarray}
\label{induzido}
\bar{\Delta}_{aa} (k, \sigma) \equiv \langle a_{k \sigma} a_{-k -\sigma} \rangle .
\end{eqnarray}
This  anomalous correlation function, as we will show below, turns out to be finite  even in  the absence of interactions in the $a$-band, due to the influence of hybridization and/or inter-band interactions.

We use the equation of motion method to find the relevant Green's functions and use the fluctuation-dissipation theorem to obtain from them the anomalous correlation functions above. Since we want to calculate the chemical potential, we also need to obtain the correlation functions  $\langle a^{\dagger}_{k \sigma} a_{k \sigma} \rangle$ and $\langle b{\dagger}_{k \sigma} b_{k \sigma} \rangle$ that yield the average number of particles in each band.

\section{\label{contas} Calculations}

In this section, we calculate the Green's functions necessary to find the intra-band and inter-band order parameters, as well as, the occupation numbers in the $a$ and $b$ bands. 

The equation of motion for the anomalous Green's function $\langle \langle b_{-k -\sigma}^{\dag} | a_{k \sigma}^{\dag} \rangle \rangle$ is given by,
\begin{eqnarray}
&& \left( \omega + \epsilon_{-k}^{b}\right) \langle \langle b_{-k -\sigma}^{\dag}| a_{k \sigma}^{\dag} \rangle \rangle +\Delta_{ab}^{*} \langle \langle a_{k \sigma} | a_{k \sigma}^{\dag} \rangle \rangle   \nonumber\\
&& \;\;\;\;\;\;\;\;+\Delta_{bb}^{*} \langle \langle b_{-k \sigma}^{\dag}| a_{-k \sigma}^{\dag} \rangle \rangle  + V_{-k} \langle \langle a_{-k -\sigma}^{\dag}| a_{k \sigma}^{\dag} \rangle \rangle = 0. \nonumber\\
\end{eqnarray}
This generates two new Green's functions for which we write the equations of motion,
\begin{eqnarray}
&&\left( \omega - \epsilon_{k}^{a}\right) \langle \langle a_{k \sigma} | a_{k \sigma}^{\dag} \rangle \rangle   \nonumber\\
&& \;\;\;\;\; + \Delta_{ab} \langle \langle b_{-k -\sigma}^{\dag} | a_{k \sigma}^{\dag} \rangle \rangle - V_{k} \langle \langle b_{k \sigma} | a_{k \sigma}^{\dag} \rangle \rangle = 1
\end{eqnarray}
and
\begin{eqnarray}
&& \left( \omega - \epsilon_{k}^{b}\right) \langle \langle b_{k \sigma} | a_{k \sigma}^{\dag} \rangle \rangle - \Delta_{ab} \langle \langle a_{-k -\sigma}^{\dag} | a_{k \sigma}^{\dag} \rangle \rangle   \nonumber\\
&& \;\;\;\;\;\;\;\;\;\;\;\;\;  + \Delta_{bb}(k, \sigma) \langle \langle b_{-k -\sigma}^{\dag} | a_{k \sigma}^{\dag} \rangle \rangle - V_{k}^{*} \langle \langle a_{k \sigma} | a_{k \sigma}^{\dag} \rangle \rangle = 0. \nonumber\\
\end{eqnarray}
Finally, we find the last Green's function namely, $\langle \langle a_{-k -\sigma}^{\dag} | a_{k \sigma}^{\dag} \rangle \rangle$, that closes the set of equations,
\begin{eqnarray}
&&\left( \omega + \epsilon_{-k}^{a} \right) \langle \langle a_{-k -\sigma}^{\dag} | a_{k \sigma}^{\dag} \rangle \rangle  \nonumber\\
&& \;\;\;\;\;\;\;\;\; - \Delta_{ab}^{*} \langle \langle b_{k \sigma} | a_{k \sigma}^{\dag} \rangle \rangle + V_{-k}^{*} \langle \langle b_{-k -\sigma}^{\dag} | a_{k \sigma}^{\dag} \rangle \rangle.
\end{eqnarray}
It is helpful to write these four equations in matrix form,
\begin{equation}
\label{eq14}
\mathbf{D} .
\left( \begin{array}{c} x_{1} = \langle \langle a_{k \sigma} | a_{k \sigma}^{\dag} \rangle \rangle\\y_{1} = \langle \langle b_{-k -\sigma}^{\dag} | a_{k \sigma}^{\dag} \rangle \rangle\\z_{1} = \langle \langle b_{k \sigma} | a_{k \sigma}^{\dag} \rangle \rangle\\u_{1} =\langle \langle a_{-k -\sigma}^{\dag} | a_{k \sigma}^{\dag} \rangle \rangle\\
\end{array}\right)=
\left( \begin{array}{c} 1\\0\\0\\0\\
\end{array}\right)
\end{equation}
where,
\begin{equation}
\label{matriz}
\mathbf{D} =\left( \begin{array}{cccc}
\left(\omega-\epsilon^{a}_{k}\right) & \Delta_{ab} & -V_{k} & 0\\
\Delta_{ab}^{*} & \left(\omega+\epsilon^{b}_{-k}\right) & \Delta_{bb}^{*} & V_{-k}\\
-V^{*}_{k} & \Delta_{bb} & \left(\omega - \epsilon^{b}_{k}\right) & - \Delta_{ab} \\
0 & V^{*}_{-k} & -\Delta_{ab}^{*} &  \left(\omega+\epsilon^{a}_{-k}\right)\\
\end{array}\right).
\end{equation}
Notice that, from this system of equations we cannot compute all the correlation functions initially desired. We also need to solve another closed set of equations that is obtained when we calculate the equation of motion for the anomalous Green's function $\langle\langle b_{-k -\sigma} | b_{k \sigma} \rangle\rangle$. This new set of equations is given by,

\begin{equation}\label{eq16}
\mathbf{D} .
\left( \begin{array}{c} x_{2} = \langle \langle a_{k \sigma} | b_{k \sigma}^{\dag} \rangle \rangle\\y_{2} = \langle \langle b_{-k -\sigma}^{\dag} | b_{k \sigma}^{\dag} \rangle \rangle\\z_{2} = \langle \langle b_{k \sigma} | b_{k \sigma}^{\dag} \rangle \rangle\\u_{2} =\langle \langle a_{-k -\sigma}^{\dag} | b_{k \sigma}^{\dag} \rangle \rangle\\
\end{array}\right)=
\left( \begin{array}{c} 0\\0\\1\\0\\
\end{array}\right).
\end{equation}

In the next subsections, we obtain the relevant correlation functions and the energy of the excitations.

\subsection{Excitation Energies}

The excitation energies of the system are given by the poles of the Green's functions. These poles are obtained from the equation $\det (\mathbf{D}) = 0$ given by,
\begin{eqnarray}
\omega^{4} - 2 A_{k} \omega^{2} + B_{k}=0
\end{eqnarray}
with
\begin{eqnarray}
\label{bk}
&&A_{k} = \frac{\epsilon_{k}^{a 2} + \epsilon_{k}^{b 2} + |\Delta_{bb}|^{2}}{2} +  |\Delta_{ab}|^{2} +  |V_{k}|^{2} \nonumber\\
&& B_{k} = \left( \epsilon_{k}^{a} \epsilon_{k}^{b} -  |V_{k}|^{2} \!+\! |\Delta_{ab}|^{2} \right)^{2}\!\!  + 4 |V_{k}|^{2}|\Delta_{ab}|^{2} \!+\! \epsilon_{k}^{a 2} |\Delta_{bb}|^2  \nonumber \\
\end{eqnarray}
where we have used the antisymmetric property of the hybridization, $V_{-k} = - V_{k}$  and that the band energies are symmetric, i.e., $\epsilon_{-k}^{a,b} = \epsilon_{k}^{a,b}$.

We assume without loss of generality that the order parameters $\Delta_{ab}$ and $\Delta_{bb}$ are real. Since the hybridization $V_k$ has to be purely imaginary to preserve time reversal symmetry, a term $-2 \epsilon_{k}^{a} \Re e \left( \Delta_{ab} V_{k} \Delta_{bb} \right)$ that  appears in $B_k$ turns out to be identically zero and has not been written above.

Notice that when $B_{k} = 0$ we find zero energy solutions:
\begin{eqnarray}
\omega^{2} \left( \omega^{2} - 2 A_{k} \right) = 0
\end{eqnarray}
These zero energy modes are associated with topological transitions in the system~\cite{mucio3, mucio2} as we will see further on in the text. 

Finally, the excitation energies are given by,
\begin{eqnarray}
\label{exc}
&& \omega_{1} = \sqrt{A_{k} + \sqrt{A_{k}^{2} - B_{k}}} = - \omega_{3} \nonumber\\
&& \omega_{2} = \sqrt{A_{k} - \sqrt{A_{k}^{2} - B_{k}}} = - \omega_{4}.
\end{eqnarray}

\section{Solution of the problem}

The inter-band order parameter is defined by Eq.~\ref{inter}. From the closed set of equations, Eqs.~\ref{eq14}, we can obtain the quantity $y_{1}$ and using the fluctuation-dissipation theorem we get the first {\it gap} equation,
\begin{eqnarray}
\label{deltaab}
&&\;\;\;\;\;\;  \frac{1}{g_{ab}} =  \frac{1}{4 \pi} \sum_{k \sigma}\left[ \frac{ \left( \omega_{1} \tanh (\beta \omega_{1}/2) \!-\! \omega_{2} \tanh (\beta \omega_{2}/2)\right)}{\omega_{1}^{2} \!-\! \omega_{2}^{2}} \right. \nonumber\\
&&  -  \frac{ \left( \epsilon_{k}^{a} \epsilon_{k}^{b} +|V_{k}|^{2}+\Delta_{ab}^{2}\right)}{\omega_{1}^{2} +\omega_{2}^{2}} \left. \left( \frac{\tanh (\beta \omega_{1}/2)}{\omega_{1}}\! -\! \frac{\tanh (\beta \omega_{2}/2)}{\omega_{2}}\right)  \right]. \nonumber\\
\end{eqnarray}

At $T = 0$ K, this becomes
\begin{eqnarray}
\label{eqinter}
\frac{1}{g_{ab}} \!= \!\frac{1}{4 \pi}  \sum_{k \sigma}\left[ \frac{1}{\omega_{1} \!+\! \omega_{2}}\left( 1 \!+\!\frac{ \epsilon_{k}^{a} \epsilon_{k}^{b} \!+\! |V_{k}|^{2} \!+\! \Delta_{ab}^{2}}{\omega_{1} \omega_{2}} \right)  \right].
\end{eqnarray}

For the intra-band {\it gap} term we get,
\begin{eqnarray}
\label{eqintra}
&& \frac{1}{g_{bb}} =  \frac{1}{4 \pi} \sum_{k \sigma} \left[ \frac{\left( \omega_{1} \tanh (\beta \omega_{1}/2)  \omega_{2} \tanh (\beta \omega_{2}/2)\right)}{\omega_{1}^{2} \!-\! \omega_{2}^{2}}  \right. \nonumber\\
&& \;\left. -  \frac{ \epsilon_{k}^{a 2}  }{\omega_{1}^{2} \!-\! \omega_{2}^{2}}\! \left( \frac{\tanh (\beta \omega_{1}/2)}{\omega_{1}}\! -\! \frac{\tanh (\beta \omega_{2}/2)}{\omega_{2}}\right)\!  \right] . \nonumber\\
\end{eqnarray}
At $T = 0$ K, this is given by,
\begin{eqnarray}
\label{deltabbtotal}
&&\frac{1}{g_{bb}}  =  \frac{1}{4 \pi}\sum_{k \sigma} \left[ \frac{1}{\omega_{1} + \omega_{2}}\left(1  + \frac{\epsilon_{k}^{a 2} }{\omega_{1} \omega_{2}}  \right)\right]. 
\end{eqnarray}
Linear terms in $V_k$ do not appear in these equations due to the assumptions that both $\Delta_{ab}$ and $\Delta_{bb}$ are real and that the system has time reversal symmetry in the absence of superconductivity.

The above equations involve the order parameters $\Delta_{ab}$, $\Delta_{bb}$ and the chemical potential $\mu$ through the dispersions of the bands. A full solution to the problem requires a self-consistent solution of a system of equations involving the three variables, $\Delta_{ab}$, $\Delta_{bb}$ and $\mu$. The equation for the chemical potential is obtained from the conservation of the total  number of particles $n$,
\begin{eqnarray}
&&n = n_{a} + n_{b} \nonumber\\
&& n = \sum_{k \sigma} \left(\langle a_{k \sigma}^{\dag} a_{k \sigma} \rangle +  \langle b_{k \sigma}^{\dag} b_{k \sigma} \rangle \right).
\end{eqnarray}
The quantities $\langle a_{k \sigma}^{\dag} a_{k \sigma} \rangle$ and  $\langle b_{k \sigma}^{\dag} b_{k \sigma} \rangle$ are obtained from the associated Green's functions ($x_1$ and $z_2$, in Eqs. (\ref{eq14}) and (\ref{eq16}), respectively).

The total number of particle is given by, 
\begin{eqnarray}
n = \frac{1}{(2 \pi)^{3}} \left[ \frac{4 \pi}{3} \left(k_{F, a}^{3} +k_{F, b}^{3} \right)\right],
\end{eqnarray}
where $k_{F, a}$ and $k_{F, b}$ are the respective wave-vectors for the $a$ and $b$ bands. We will assume that these bands are homotetic, i.e.,  
\begin{eqnarray}
&&\epsilon_{k}^{a} = \epsilon_{k} - \mu \nonumber\\
&&\epsilon_{k}^{b} = \alpha \epsilon_{k} - \mu,
\end{eqnarray}
where $\epsilon_{k} = k^{2}/2m_{a}$. The quantity $\alpha = m_{a}/m_{b}$ is the ratio of the effective masses. Notice that, $k_{F, b} = k_{F, a}/ \sqrt{\alpha}$. The Fermi energy is given by, $E_{F} = k_{F, a(b)}^{2}/2m_{a(b)}$ and the total number of particles can be expressed as,  
\begin{eqnarray}
\label{ocfermi}
n = \frac{1}{(2 \pi)^{3}} \left[ \frac{4 \pi k_{F, a}^{3}}{3} \left( \frac{1 + \alpha^{3/2}}{\alpha^{3/2}} \right)\right].
\end{eqnarray}

Using Eq.~\ref{ocfermi}, we write the equation for the occupation number as,
\begin{eqnarray}
\label{occ}
&& 1 = \frac{3 \pi}{2 k_{F, a}^{3}}\left(  \frac{\alpha^{3/2}}{1 + \alpha^{3/2}} \right)\sum_{k \sigma} \left[ 2 - \frac{\epsilon_{k}^{a} + \epsilon_{k}^{b}}{\omega_{1} + \omega_{2}}  \right. \nonumber\\
&& \;\;\;\;\; \left.- \frac{\left( \epsilon_{k}^{a} + \epsilon_{k}^{b} \right) \left( \epsilon_{k}^{a}\epsilon_{k}^{b} + \Delta_{ab}^{2} - |V_{k}|^{2}\right) + \epsilon_{k}^{a} \Delta_{bb}^{2}}{\omega_{1}\omega_{2} \left( \omega_{1} + \omega_{2}\right)} \right],
\end{eqnarray}
at $T = 0$ K.

Eqs.~\ref{eqinter},~\ref{deltabbtotal} and~\ref{occ}  define our problem with intra and inter-band superconductivity. For solving them summation in $k$ is changed to integration using,
\begin{eqnarray}
\sum_{k \sigma} \longrightarrow \frac{V}{(2 \pi)^{3}} k_{F, a}^{3} \int d^{3} \tilde{k}.
\end{eqnarray}
Here and below {\it tilde} quantities mean, for wave-vectors normalization by $k_{F, a}$ and for energies normalization by $E_F$. $V$ is the volume element that we  take, without loss of generality, as $V = 1$.

We will not investigate in this work the full self-consistent problem involving the two superconducting order parameters and the chemical potential. Instead, we  consider the particular cases where we have, either intra or inter-band interactions only. 
Furthermore, we want to be able to extend our calculations to the strong coupling regime where $g_{ab}$ and $g_{bb}$ are very large. As usual, in the study of the crossover from the weak to the strong coupling regimes, we  introduce two scattering lengths $a_s$ and $a_{sb}$~\cite{randeria, pairing} which replace the coupling constants $g_{ab}$ and $g_{bb}$, respectively. Since the integrals are done for all values of $k$, they diverge and have to be regularized. We employ here the standard procedure to eliminate these ultra-violet divergences (see below).

\begin{itemize}
\item Inter-band case

In the pure inter-band case $g_{bb}$ is zero and there are only inter-band attractive interactions in the system.
We are left with two equations for $\Delta_{ab}$ and the chemical potential $\mu$ to be solved self-consistently.
These are given by,
\begin{eqnarray}
\label{critico1}
 &&\frac{1+\alpha}{4\pi^{3}} \int d^3\tilde{k} \left[ \frac{1}{ \tilde{\omega}_{1} + \tilde{\omega}_{2}} - \frac{2}{(1+ \alpha) \tilde{k}^{2}} \; \right.\nonumber\\
&&\;\;\;\;\;\;\;\;\;\;\;\; \left.+\; \frac{\left( \tilde{\Delta}_{ab}^{2} + \tilde{\epsilon}_{k}^{a} \tilde{\epsilon}_{k}^{b} +|\tilde{V}_{k}|^{2}\right)}{\tilde{\omega}_{1} \tilde{\omega}_{2}\left( \tilde{\omega}_{1} + \tilde{\omega}_{2}\right)}  \right]  =- \frac{1}{k_{F} a_{s}} ,
\end{eqnarray}
and the number equation,
\begin{eqnarray}
\label{numer1}
&& \frac{3}{16 \pi^{2}} \left( \frac{\alpha^{3/2}}{1 + \alpha^{3/2}}\right)\int d^3\tilde{k} \left[ 2 \!- \frac{\tilde{\epsilon}_{k}^{a} \!+\! \tilde{\epsilon}_{k}^{b}}{\tilde{\omega}_{1} + \tilde{\omega}_{2}}   \right. \nonumber\\
&& \;\;\; \left. - \frac{\left( \tilde{\epsilon}_{k}^{a} \!+\! \tilde{\epsilon}_{k}^{b}\right) \left(\tilde{\epsilon}_{k}^{a} \tilde{\epsilon}_{k}^{b} \!-\! |\tilde{V}_{k}|^{2} \!+\! |\tilde{\Delta}_{ab}|^{2} \right) }{\tilde{\omega}_{1} \tilde{\omega}_{2} \left( \tilde{\omega}_{1} + \tilde{\omega}_{2}\right)} \right] = 1. \nonumber\\
\end{eqnarray}

We used the definition of the scattering length for the low energy limit of the two-body problem in the vacuum~\cite{randeria},  
\begin{eqnarray}
\label{reg}
\frac{1}{g_{ab}} = - \frac{m^{*}}{4 \pi a_{s}} +  \int d^3k \left[ f_{1} \left( k\right) - f_{1} \left( k \rightarrow \infty \right)\right]
\end{eqnarray}
where $a_{s}$ is the s-wave scattering length, $m^*= \alpha/(1+\alpha) m_a$ and $\alpha=m_a/m_b$ is the ratio of the effective masses of the $a$ and $b$ quasi-particles. 
The function $f_1(k)$ in this case  is given by, 
\begin{eqnarray}
f_{1} \left( k\right) \!=\!\left[ \frac{1}{\omega_{1}\! +\! \omega_{2}}\!  +\! \frac{ \left(\! \epsilon_{k}^{a} \epsilon_{k}^{b} \!+\! |V_{k}|^{2} \!+\! |\Delta_{ab}|^{2}\right)}{\omega_{1} \omega_{2} (\omega_{1} + \omega_{2})}  \right] . \nonumber
\end{eqnarray}
Since the integral extends to infinity the subtraction of the last term on the right hand side of Eq. (\ref{reg}) regularizes the ultra-violet divergence in this expression. 

\item Intra-band case

In this case, we take $g_{ab}=0$ and the gap equation is given by,
\begin{eqnarray}
\label{critico2}
\;\;\;\;\;\;\; -\frac{1}{k_{F} a_{sb}} \!=\! \frac{\alpha}{4\pi^{3}} \int \! d^3\tilde{k} \left[ \frac{1}{\tilde{\omega}_{1} \!+\! \tilde{\omega}_{2}}  \left( 1 \! +\! \frac{ \tilde{\epsilon}_{k}^{a2}  \tilde{\Delta}_{bb} } { \tilde{\omega}_{1} \tilde{\omega}_{2}} \right) \! - \! \frac{ 1} { \alpha \tilde{k}^{2} } \right], \nonumber\\
\end{eqnarray}
where we replaced $g_{bb}$ by the  scattering length. 
The occupation number equation is now given by,
\begin{eqnarray}
\label{numer}
&&\;\;\; \frac{3}{16 \pi^{2}} \left( \frac{\alpha^{3/2}}{1 + \alpha^{3/2}}\right)\int d^3\tilde{k}  \left[ 2 - \frac{\tilde{\epsilon}_{k}^{a} \!+\! \tilde{\epsilon}_{k}^{b}}{\tilde{\omega}_{1} + \tilde{\omega}_{2}}  \right. \nonumber\\
&&  \left.   +\! \frac{\tilde{\epsilon}_{k}^{a} |\tilde{\Delta}_{bb}|^{2}}{\tilde{\omega}_{1} \tilde{\omega}_{2} \left( \tilde{\omega}_{1}\! +\! \tilde{\omega}_{2}\right)} \! -\! \frac{\left( \tilde{\epsilon}_{k}^{a} \!+\! \tilde{\epsilon}_{k}^{b}\right) \left(\tilde{\epsilon}_{k}^{a} \tilde{\epsilon}_{k}^{b} \!-\! |\tilde{V}_{k}|^{2}  \right)  }{\tilde{\omega}_{1} \tilde{\omega}_{2} \left( \tilde{\omega}_{1} + \tilde{\omega}_{2}\right)} \right] = 1. 
\end{eqnarray}
\end{itemize}

Notice that in each of the cases above the excitation energies $\tilde{\omega}_1$ and $\tilde{\omega}_2$ are different. They are obtained from Eqs.~\ref{exc} making either $\Delta_{ab}=0$ or $\Delta_{bb}=0$, as appropriate. As before $tilde$ quantities mean, for energies normalization by the Fermi energy $E_{F}$, for wave-vectors by $k_{F,a}$.

\section{Numerical Solution}

\subsection{Inter-band case -- $g_{bb} = 0$}

\begin{figure}[!htb]
\centering
\subfigure[]{
\includegraphics[width=1\linewidth]{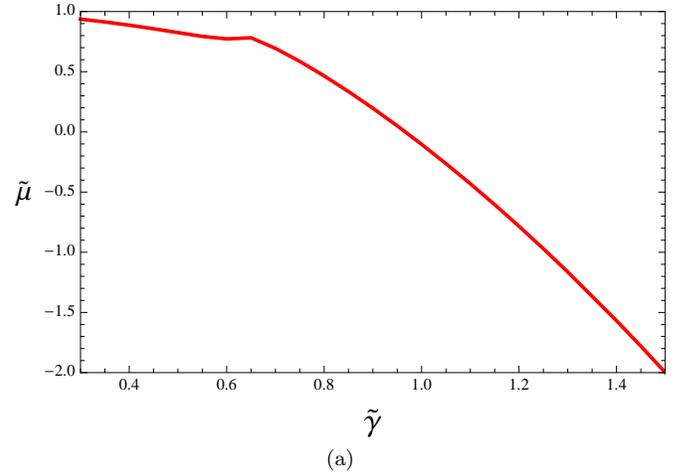}
}
\subfigure[]{
\includegraphics[width=1\linewidth]{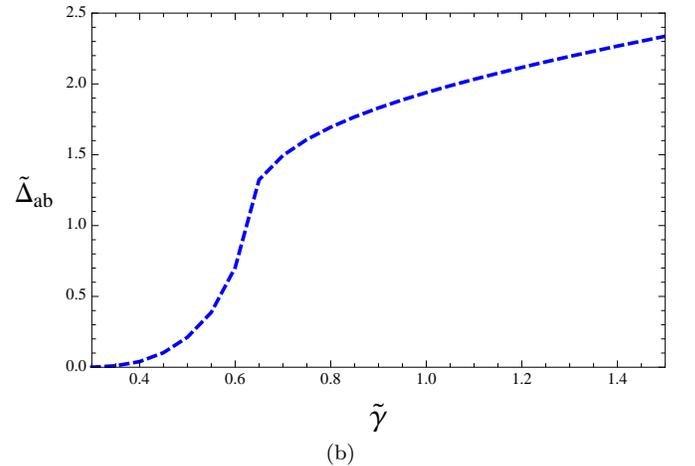}
}
\caption{(Color online) The normalized chemical potential $\tilde{\mu}$ and the   inter-band superconducting order parameter  $\tilde{\Delta}_{ab}$ as  functions of the strength of the hybridization for fixed $1/k_{F}a_{s} = - 0.5 $. This value for the coupling is typical of the weak coupling BCS regime.  We used $\alpha = 0.1$ for the ratio of the effective masses and $\beta=0.1$ for the {\it anisotropy} parameter.  Notice that the hybridization induces a BCS-BEC crossover. }
\label{fig1}
\end{figure}

In this section we solve self-consistently Eqs.~\ref{critico1} and~\ref{numer1} for the inter-band superconducting order parameter $\Delta_{ab}$ and the chemical potential $\mu$. We consider a three dimensional system and take for the hybridization the form $V(\vec{k})=i\gamma\left(k_x + k_y + \beta k_z \right)$. For $\beta<1$ we want to describe a tetragonal system where the hybridization is smaller between planes.  Since we are here interested in studying the effect of hybridization on superconductivity, we take the quantity $1/k_{F} a_{s}$ negative and small, such that the system is in the weak coupling BCS regime.  The results for $\Delta_{ab}$ and $\mu$ are shown in Fig. (\ref{fig1}). They correspond to fixed $1/k_{F} a_{s}=-0.5$ and a ratio for the masses $\alpha=0.1$. Also the parameter $\beta=0.1$ in $V_k$.
We see from Fig. (\ref{fig1}) that the normalized chemical potential $\tilde{\mu}$ decreases as the strength of the antisymmetric hybridization $\tilde{\gamma}=\gamma/v_F$ increases. This decrease is accompanied by an increase of the inter-band superconducting order parameter $\tilde{\Delta}_{ab}$.
There are at least three very interesting points to be noticed in these figures. First, the enhancement of superconductivity by antisymmetric hybridization which had already been noticed~\cite{mucio2}. Second, the combined behavior of the chemical potential decreasing $\tilde{\gamma}$ and the concomitant increase in $\tilde{\Delta}_{ab}$ is a clear signature of a BCS-BEC crossover induced in this case by an increase in hybridization~\cite{mucio2}.  Finally, notice the discontinuous behavior of $\tilde{\mu}$ and $\tilde{\Delta}_{ab}$ for $\tilde{\gamma} \approx 0.65$, due to a topological phase transition associated with the appearance of gapless modes in the spectrum of excitations for this value of $\tilde{\gamma}$. This is due to a vanishing of the quantity $B_k$ in Eq. (\ref{bk}).
For $\Delta_{bb}=0$, the condition $B_k=0$ implies $V_k=0$ and $\epsilon_k^a \epsilon_k^b + \Delta_{ab}^2=0$. The intersection of these two surfaces give the line of zero energy excitations associated with the topological transition. This transition is very sensitive to the ratio of the effective masses and for $\alpha=0.5$ it has disappeared for the range of $\tilde{\gamma}$ in Fig.~\ref{fig1}.

\subsection{Intra-band case -- $g_{ab} = 0$}
\begin{figure}[!htb]
\centering
\subfigure[]{
\includegraphics[width=1\linewidth]{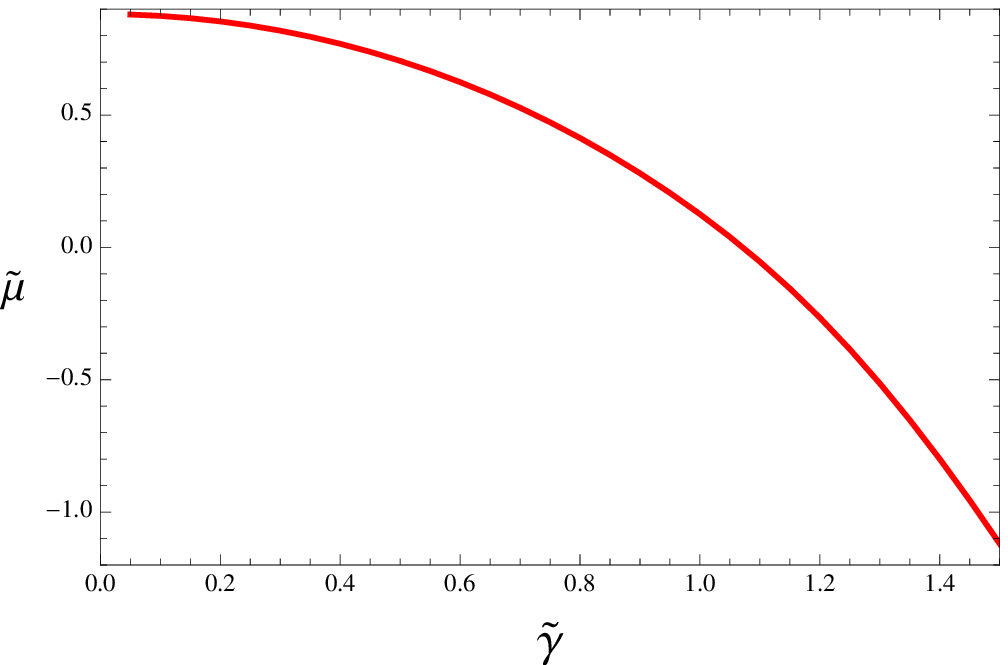}
}
\subfigure[]{
\includegraphics[width=1\linewidth]{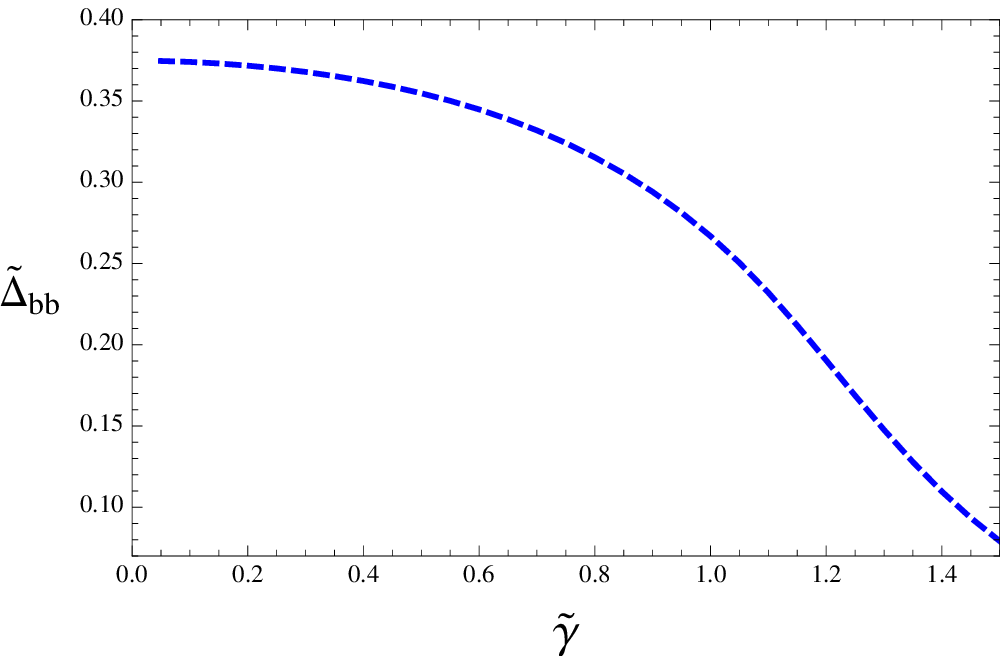}
}
\caption{(Color online) The normalized chemical potential $\tilde{\mu}$ and the   intra-band superconducting order parameter  $\tilde{\Delta}_{bb}$ as  functions of the strength of the hybridization for fixed $1/k_{F}a_{sb} = - 0.5 $, typical of the weak coupling BCS regime.  We used $\alpha = 0.1$ for the ratio of the effective masses and $\beta=0.1$ for the {\it anisotropy} parameter.   } 
\label{fig2}
\end{figure}

Here we discuss  the  pure intra-band case, where $g_{ab} = 0$  in Eq.~\ref{hamiltonian}. In this case superconductivity is characterized by the intra-band order parameter $\Delta_{bb}$, defined in Eq.~\ref{parintra}. Taking $\Delta_{ab}=0$  we are left with a system of two equations to be solved self-consistently. As in the previous inter-band case, we assume for the antisymmetric hybridization the form $V(\vec{k})=i\gamma\left(k_x + k_y + \beta k_z \right)$ where $V_{k}$ is purely imaginary to preserve time reversal symmetry.
The self-consistent solutions for the order parameter $\Delta_{bb}$ and the chemical potential in terms of $\gamma / v_{F} = \tilde{\gamma}$ are shown in  Fig.~\ref{fig2}  of this section.  
Here we recover the usual deleterious behavior of the hybridization on intra-band superconductivity~\cite{igor, mucio}. We can see that $\tilde{\Delta}_{bb}$ differently from $\tilde{\Delta}_{ab}$ decreases with increasing $\tilde{\gamma}$ eventually vanishing at a superconductor quantum critical point.

\section{Induced Superconductivity}

An interesting result of our investigation is  that,  in spite of the absence of  interactions in the $a$-band, the anomalous correlation function $\Delta_{aa}(k)$ may assume finite values, signaling the presence of superconducting correlations in this non-interacting band. The anomalous correlation function $\Delta_{aa}(k)=\langle a_{k \sigma} a_{-k -\sigma}\rangle$ can be easily calculated from the associated Green's function ($u_{1}$) and we get,

\begin{eqnarray}
\label{eqindu}
&&\Delta_{aa}(k) = - \frac{1}{4 \pi}  \left[ \frac{2 \epsilon_{k}^{b} V_{k} \Delta_{ab} - \Delta_{bb} \left(|V_{k}|^{2} - \Delta_{ab}^{2} \right)}{\omega_{1}^{2} - \omega_{2}^{2}}  \right. \times \nonumber\\
&&\;\;\;\;\;\;\;\;\;\;\; \times \left. \left( \frac{\tanh (\beta \omega_{1}/2)}{\omega_{1}} -  \frac{\tanh (\beta \omega_{2}/2)}{\omega_{2}}\right) \right].
\end{eqnarray}

At $T = 0$ K, this becomes,
\begin{eqnarray}
\label{induced}
\label{daltaa}
\Delta_{aa}(k) =  \frac{1}{4 \pi}  \left[ \frac{2 \epsilon_{k}^{b} V_{k} \Delta_{ab} - \Delta_{bb} \left(|V_{k}|^{2} - \Delta_{ab}^{2} \right)}{\omega_{1} \omega_{2} \left(\omega_{1} + \omega_{2}\right)}    \right].
\end{eqnarray}
This induced anomalous correlation function  in the $a$-band arises due to the influence of hybridization and/or inter-band interactions. A linear term in $V_k$ now appears, indicating that  $\Delta_{aa}(k)$ can be complex. This correlation function is a sum of  antisymmetric and symmetric contributions. Notice that $\epsilon_k^b$, $\omega_1$ and $\omega_2$ are even functions of $k$.
The antisymmetric term proportional to $V_k$ corresponds to induced $p$-wave anomalous correlations due to the odd-parity of $V_k$. It preserves time reversal invariance since $\Delta_{ab}$ is real and $V_k$ is purely imaginary. This term corresponds to the component $m=0$ of the $l=1$ $p$-wave state.

Eq. (\ref{daltaa}) for the induced anomalous correlation function $\Delta_{aa}$ is very interesting, and deserves special attention. We notice that there are three possible situations involving the relative magnitudes of the order parameters, that will be discussed now.

\begin{enumerate}

\item The first situation is met when the second term in the right side of the above equation can be neglected, i.e., when $2 \epsilon_{k}^{b} |V_{k}| \Delta_{ab} \gg \Delta_{bb} \left(|V_{k}|^{2} - \Delta_{ab}^{2} \right)$ with the $k$-dependent quantities calculated at $k_F^a$. Observe that this is naturally true when $\Delta_{bb}$ is very small (compared to $\Delta_{ab}$) and also when $|V_{k}| \sim \Delta_{ab}$, regardless of the value of $\Delta_{bb}$. Then we can write,
\begin{eqnarray}
\label{dalta1}
\Delta_{aa}(k) \approx  \frac{1}{2 \pi}  \left[ \frac{\epsilon_{k}^{b} V_{k} \Delta_{ab}}{\omega_{1} \omega_{2} \left(\omega_{1} + \omega_{2}\right)} \right].
\end{eqnarray}
Since the parity of $\Delta_{aa}$ follows that of $V_{k}$ which is antisymmetric,  $\Delta_{aa}(k)$ has odd-parity in $k$-space. Then the superconducting fluctuations induced in the $a$-band have a $p$-wave character.

\item The second one happens in the opposite limit, that is  $\Delta_{bb} \left(|V_{k}|^{2} - \Delta_{ab}^{2}\right)  \gg 2 \epsilon_{k}^{b} |V_{k}| \Delta_{ab} $, which can be obtained, for instance, for a vanishingly small $V_{k}$, resulting in,
\begin{eqnarray}
\label{dalta2}
\Delta_{aa}(k) \approx  \frac{1}{4 \pi}  \left[ \frac{\Delta_{bb} \Delta_{ab}^2 }{\omega_{1} \omega_{2} \left(\omega_{1} + \omega_{2}\right)} \right].
\end{eqnarray}
See that in this case $\Delta_{aa}$ is completely symmetric.

\item The third case  is obtained when $\Delta_{ab} \approx 0$,
\begin{eqnarray}
\label{dalta}
\Delta_{aa}(k) =  \frac{1}{4 \pi}  \left[ \frac{ - \Delta_{bb} |V_{k}|^{2}} 
{\omega_{1} \omega_{2} \left(\omega_{1} + \omega_{2}\right)}    \right].
\end{eqnarray}
This corresponds to the more conventional induced superconductivity, as in the proximity effect. It is a second order effect in the hybridization.
In the last two cases the induced superconductivity is clearly $s$-wave.
\end{enumerate}

\section{Majorana fermions}

There is nowadays a great interest in the study of Majorana fermions as possible candidates for use as q-bits in quantum computers. 
In an exciting paper~\cite{kitaev}, Kitaev has shown that one-dimensional $p$-wave superconductors can exhibit a non-trivial topological phase with Majorana fermions at the ends of the chain. A great effort is being done to implement this idea of Kitaev in an actual physical wire~\cite{sarma}. The main difficulty is of course that $p$-wave superconductors are scarce in nature and to circumvent this, many suggestions have been proposed. The most successful one combine a mixture of spin-orbit interaction and external magnetic field to induce in a semiconductor wire $p$-wave type of superconductivity~\cite{sarma}.

The results above suggest a new mechanism for inducing $p$-wave superconductivity in a wire and consequently to obtain Majorana fermions. The idea is to consider the Hamiltonian, Eq.~\ref{hamiltonian}, as an effective Hamiltonian for the following system. A bulk BCS superconductor with  a $b$-band of electrons with an  attractive interaction  responsible for the superconductivity characterized by the $s$-wave order parameter $\Delta_{bb}$. On top of this material, there is a normal wire with a non-interacting band of $a$-electrons which are coupled to the bulk substrate through an antisymmetric hybridization and  a weak attractive interaction $g_{ab}$. The latter gives rise to the inter-band pairing $\Delta_{ab}$ among the electrons of the wire and the substrate. Then, under the conditions $1$ of Section $VI$, we obtain an induced $p$-wave superconductivity in the wire. 

In order to verify if the induced superconductivity in the wire gives rise Majorana fermions at its ends, we need to write the wire Hamiltonian in terms of Majorana operators. First, we present the Hamiltonian of the wire in terms of fermions operators,
\begin{eqnarray}
\label{h2}
&&\;\;\;\;\;\;\;\;\; H_{wire} = -\sum_{j =1}^{N} \left( \mu_{\uparrow} u_{j}^{\dag} u_{j} + \mu_{\downarrow} d_{j}^{\dag} d_{j}\right)  \nonumber\\
&&- t \sum_{j= 1}^{N-1}  \left( u_{j}^{\dag} u_{j+1} \!+\! u_{j+1}^{\dag} u_{j} d_{j}^{\dag} d_{j+1} \!+\! d_{j+1}^{\dag} d_{j}\right)  \nonumber\\
&& + \sum_{j=1}^{N-1} \Delta \left(u_{j} d_{j+1} - u_{j+1} d_{j} + d_{j+1}^{\dag} u_{j}^{\dag} - d_{j}^{\dag} u_{j+1}^{\dag}\right),
\end{eqnarray}
where $u^{\dagger}$ and $d^{\dagger}$ are creation operators for spins up and down in the $a$-band of the wire, respectively. The spin dependent chemical potentials are given by,  $\mu_{\uparrow (\downarrow)} = \mu \pm h$, where $h$ an external magnetic field. The quantity $t$ is a nearest neighbor  hopping  and $\Delta=|\Delta_{ij}|$ is the induced antisymmetric pairing. The antisymmetric nature of this coupling has already been taken into account when writing the Hamiltonian in the form above.  Notice that $\Delta_{ij}= \sum_k \Delta_{aa}(k) e^{ik(r_i-r_j)}$ which is antisymmetric in real space since $\Delta_{aa}(k)  \propto V_{k}$ is the induced $p$-wave superconductivity in the wire due to the antisymmetric mixing between the electrons in the wire and the bulk ($V_{-k}=-V_{k}$). This pairing interaction couples electrons of opposite spins. It corresponds to the $m_{l}=0$ component of the $l=1$ $p$-wave state. This differently from the $m = \pm 1$ components  does not break time reversal symmetry. This is broken in the Hamiltonian above  by the external magnetic field $h$.

\begin{figure}[!htb]
\centering
\includegraphics[width=1\linewidth]{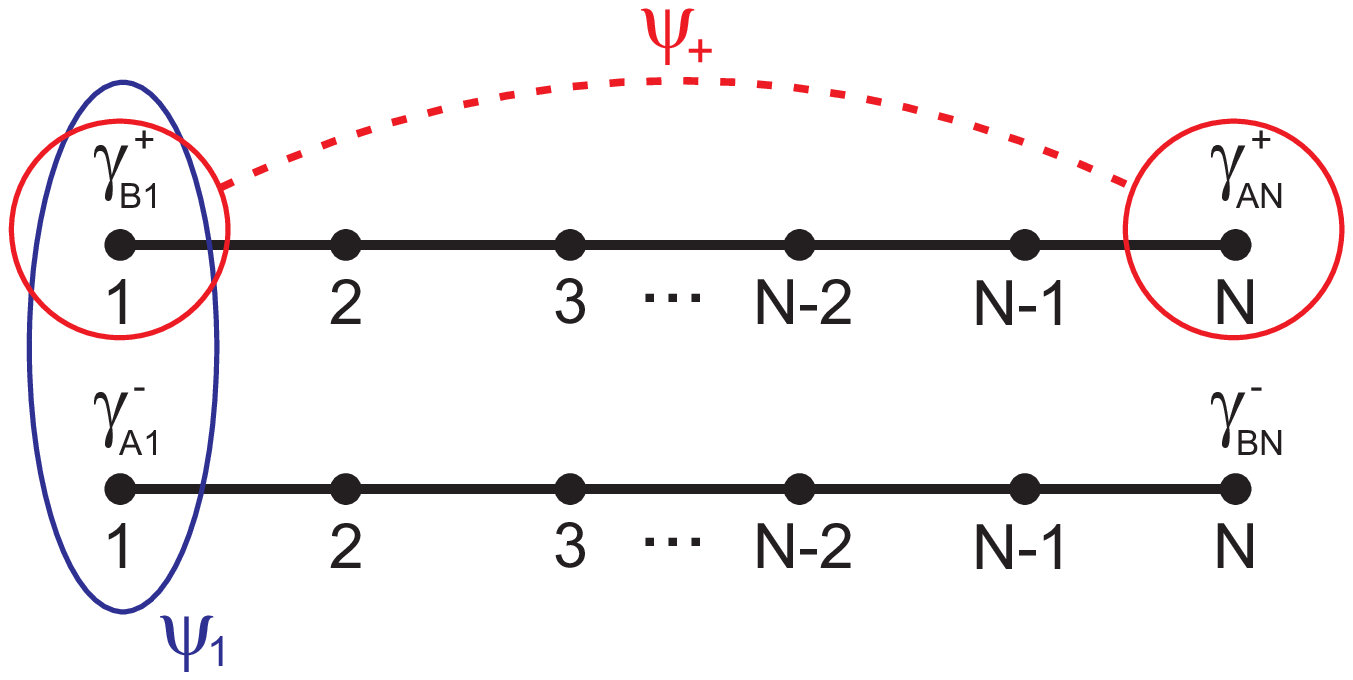}
\caption{(Color online) Schematic figure showing the two Kitaev chains, $+$ and $-$, associated with the wire. There are four Majorana states at the ends of the wire. The \textit{local} and \textit{non-local} combinations of the Majoranas in fermion states are also shown.} 
\label{fig3}
\end{figure}

We can rewrite the fermion creation and annihilation operators of Hamiltonian Eq.~\ref{h2} in terms of Majorana fermions operators. These are given by the following definitions,
\begin{eqnarray}
&&u_{j} = \frac{1}{\sqrt{2}} \left( \alpha_{B j} + i \alpha_{A j}\right) \nonumber\\
&&d_{j} = \frac{1}{\sqrt{2}} \left( \beta_{B j} + i \beta_{A j}\right)
\end{eqnarray}
and similar equations for their complex conjugates, noticing that
the Majorana particles are their own antiparticles, i.e., $\alpha_{A j}^{\dag} =\alpha_{A j} $, $\alpha_{B j}^{\dag} =\alpha_{B j}$, $\beta_{A j}^{\dag} =\beta_{A j}$ and $\beta_{B j}^{\dag} =\beta_{B j} $. In the Majorana basis the Hamiltonian Eq.~\ref{h2} becomes, 
\begin{eqnarray}
\label{h3}
&&\;\;\;\;\; H_{wire} = - \frac{\mu}{2} - i \sum_{j = 1}^{N} \left( \mu_{\uparrow} \alpha_{B j} \alpha_{A j} + \mu_{\downarrow} \beta_{B j} \beta_{A j}\right)  \nonumber\\
&&- i \sum_{j = 1}^{N-1} t \left(\alpha_{B j} \alpha_{A j\!+\!1} \!-\! \alpha_{Aj} \alpha_{B j\!+\!1} \!+\! \beta_{B j} \beta_{A j\!+\!1} \!-\! \beta_{A j} \beta_{B j\!+\!1}\right)  \nonumber\\
&&+ i \sum_{j = 1}^{N-1} \Delta \left( \alpha_{A j} \beta_{B j\!+\!1} \!+\! \alpha_{B j} \beta_{A j\!+\!1} \!-\! \alpha_{A j\!+\!1} \beta_{B j} \!-\! \alpha_{B j\!+\!1} \beta_{A j}\right). \nonumber\\
\end{eqnarray}
We introduce new hybrid Majorana operators, like in Ref.~\cite{mucio3} to get,
\begin{eqnarray}
&&\gamma_{A j}^{\pm} = \alpha_{A j} \pm \beta_{A j} \nonumber\\
&&\gamma_{B j}^{\pm} = \alpha_{B j} \pm \beta_{B j},
\end{eqnarray}
and rewrite Eq.~\ref{h3} in term of these new operators. We obtain,
\begin{eqnarray}
\label{hwire}
&&H_{wire} = - \frac{\mu}{2}  \nonumber\\
&&-  \frac{i\mu}{2} \sum_{j = 1}^{N}\left[ \left( \gamma_{Bj}^{+} \gamma_{A j}^{+} \!+\! \gamma_{B j}^{-} \gamma_{A j}^{-}\right) \!-\! \frac{i h }{2}  \left( \gamma_{Bj}^{+} \gamma_{A j}^{-} \!+\! \gamma_{B j}^{-} \gamma_{A j}^{+}\right)\right]  \nonumber\\
&&- \frac{ i(t - \Delta)}{2} \sum_{j = 1}^{N} \left( \gamma_{Bj}^{+} \gamma_{A j+1}^{+} - \gamma_{A j}^{-} \gamma_{B j+1}^{-}\right)  \nonumber\\
&&- \frac{i (t + \Delta)}{2} \sum_{j = 1}^{N} \left( \gamma_{Bj}^{-} \gamma_{A j+1}^{-} - \gamma_{A j}^{+} \gamma_{B j+1}^{+}\right).
\end{eqnarray}
This equation shows that the magnetic field $h$ couples the operators $\gamma^{\pm}$. When this field is zero, the Hamiltonian, Eq.~\ref{hwire}, reduces to that of two decoupled Kitaev chains. It is easy to see that for $h=0$, this model has a topological superconducting phase for $|\mu/2t| < 1$ as the Kitaev model~\cite{kitaev}.
Let us consider $\mu = h = 0$ and $\Delta = t$, such that the system is in the topological phase. In this case the Hamiltonian of the wire is given by, 
\begin{eqnarray}
&&H_{wire} = -i t  \sum_{j = 1}^{N} \left( \gamma_{Bj}^{-} \gamma_{A j+1}^{-} - \gamma_{A j}^{+} \gamma_{B j+1}^{+}\right).
\end{eqnarray}
Notice that the hybrid Majorana operators $\gamma_{A1}^{-}$ and $\gamma_{B 1}^{+}$ do not enter in this Hamiltonian, so that, there are two unpaired hybrid Majorana fermions in the left end of the chain. The same occurs for $\gamma_{A N}^{+}$ and $\gamma_{B N}^{-}$ on the right end of the chain. We can combine these four Majorana fermions as two {\it ordinary} fermions, $\psi_{1} = \left(\gamma_{B1}^{+} + i \gamma_{A1}^{-} \right)/2$ and $\psi_{N} = \left( \gamma_{BN}^{-} - i \gamma_{AN}^{+}\right)/2$. Also we can combine the Majoranas at different ends of the wire, such that, $\psi_{+} = \left( \gamma_{B1}^{+} + i \gamma_{AN}^{+}\right)/2$ and $\psi_{-}=\left(\gamma_{BN}^{-} - i \gamma_{A1}^{-} \right)/2$. Fig.~\ref{fig3} shows schematically these combinations.
Next, we define local pseudo-spin operators as in Ref.~\cite{chines},
\begin{eqnarray}
{\mathbf s}_{1(N)} = \frac{1}{2} \gamma_{1(N) i} {\mathbf \sigma}_{i j} \gamma_{1(N) j},
\end{eqnarray}
where the components of ${\mathbf \sigma}$ are the Pauli matrices, $i$ and $j$ can be $+$ and $-$, and  $\gamma_{1+} \equiv \gamma_{B1}^{+}$, $\gamma_{1 -} \equiv \gamma_{A1}^{-}$, $\gamma_{N+} \equiv \gamma_{AN}^{+}$ and $\gamma_{N-} \equiv \gamma_{BN}^{-}$. Using the anticommutation relations of the Majorana operators, $\{\gamma_{i +}, \gamma_{j -}\} = 0$ and $\{\gamma_{i +(-)}, \gamma_{j +(-)}\} = 2 \delta_{ij}$, we obtain  $\mathbf{s}_{1(N)}$. It turns out that both $\mathbf{s}_{1}$ and $\mathbf{s}_{N}$ have only the $y$ component different from zero. In this sense we conclude that these composite Majorana fermions behave  as Ising spins. Furthermore, we have,  $s_{1(N)y} = (i\gamma_{1(N)-}\gamma_{1(N)+})/2$, that can be rewritten as,
\begin{eqnarray}
s_{1(N)y}=\frac{1}{2} - \psi_{1(N)}^{\dag} \psi_{1(N)},
\end{eqnarray}  
where $\psi_{1(N)}$ were defined above. It is interesting to write this expression in this way because we can relate these operators to the occupation numbers of the fermion states at each end of the wire.

\begin{figure}[!htb]
\centering
\includegraphics[width=1\linewidth]{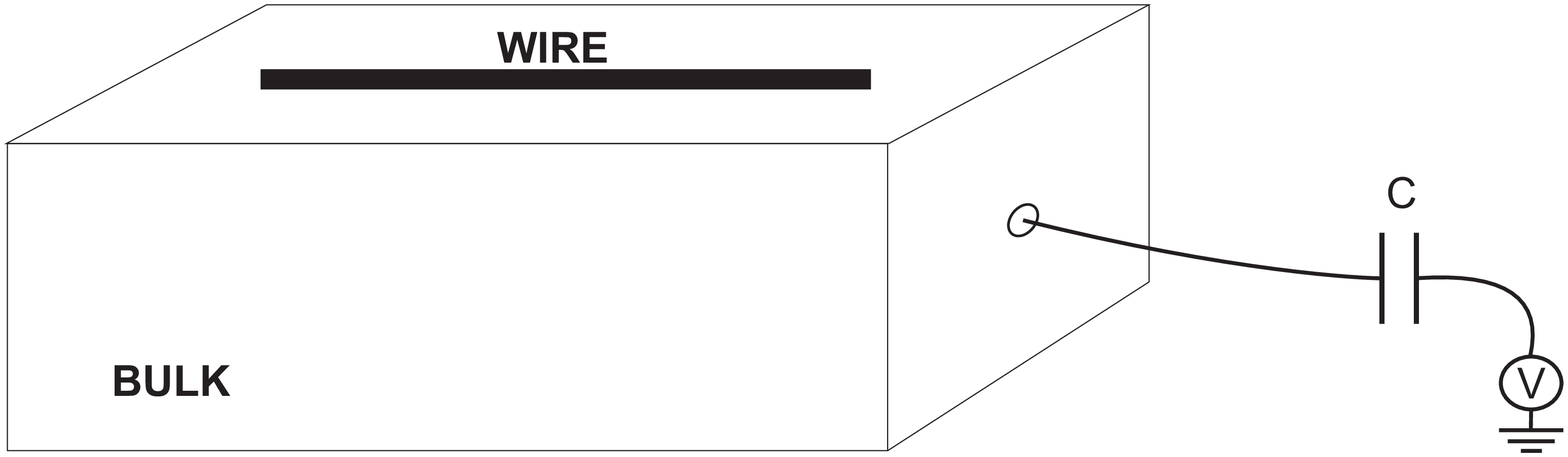}
\caption{(Color online) Schematic figure showing the bulk-wire system connected to a capacitor to modify the fermion number via a charging energy: $U(n) = Q^{2}/(2C) - Q\mathcal{V}(t)$. Here, the number of fermions is changed by variating $\mathcal{V}$~\cite{chines}.} 
\label{fig4}
\end{figure}

As in Refs.~\cite{chines}, we connect our system (superconductor bulk and wire) to a capacitor as in Fig.~\ref{fig4} to control the occupation numbers of the fermion states in the ends of the wire. There will be a total of four occupancy states: $|0\rangle_{1}|0\rangle_{N}$, $|1\rangle_{1}|1\rangle_{N}$, $|0\rangle_{1}|1\rangle_{N}$ and $|1\rangle_{1}|0\rangle_{N}$. Notice that the first two states have an even number of fermions and the last two an odd number. We can define two density matrices: $\rho_{odd}\! =\! 1/2\left( |0\rangle_{1} |1\rangle_{N \;1} \langle 0|_{N} \langle 1| + |1\rangle_{1} |0\rangle_{N \;1} \langle 1|_{N} \langle 0|  \right)$ with odd parity states and $\rho_{even} = 1/2\left( |0\rangle_{1} |0\rangle_{N \;1} \langle 0|_{N} \langle 0| + |1\rangle_{1} |1\rangle_{N \;1} \langle 1|_{N} \langle 1|  \right)$ with even parity states. We then calculate the value of the correlation function $<s_{1y}s_{Ny}>$ in these different parity states. Since, $<A>=$Tr$ \rho A$, we get,
\begin{eqnarray}
- \langle s_{1y}s_{Ny}\rangle_{odd} = \langle s_{1y}s_{1N} \rangle_{even} = \frac{1}{4}.
\end{eqnarray}
This shows that there are long range correlations between the end states of the wire, which are independent of its length. Controlling the charge of the end states we can vary the sign of these long range correlations. These highly non-trivial properties of this type of $p$-wave superconductor in its non-trivial topological phase  can certainly find useful applications, for example, in quantum computers.

\section{\label{fim}Conclusions}

In this paper, we have studied the influence of an odd-parity hybridization in the superconducting properties of a two-band system with intra (in only one of the bands)  and inter-band attractive interactions. First, we discussed the origin and conditions for antisymmetric hybridization, and showed that  it occurs in many systems of great interest as organic materials,  copper oxides and heavy fermion systems. In these systems the relevant bands consist of electrons whose angular momenta  differ by an odd number, as for $s$-$p$, $p$-$d$ and $d$-$f$-electrons, which is the required condition for antisymmetric mixing. Antisymmetric hybridization is also a crucial element for the appearance  of topological insulating phases in multi-band systems~\cite{spchain}.

Using a Green's function approach, we obtained a system of three self-consistent equations for the  intra and inter-band superconducting order parameters and the chemical potential. Our approach incorporates a scattering length  and allows to treat the cases of weak and strong inter and intra-band attractive interactions and consequently to study the crossover between the weak coupling BCS regime and the Bose-Einstein condensation (BEC) of pairs. This crossover has been intensively studied theoretically and its experimental observation was only possible with the advancement of cooling techniques for atomic gases. In these systems it is possible to control the strength of the many-body interactions, which is not feasible in condensed matter systems. Here we obtained  the remarkable result that hybridization can drive the BCS-BEC crossover even keeping the interactions fixed at values characteristic of the weak coupling regime. Since hybridization can be tuned in solid state matter by applying pressure, such as to vary the overlap of the wave-functions, this opens  a unique possibility for investigating experimentally the BCS-BEC crossover in condensed matter systems.

We have studied the induced superconductivity that appears in the non-interacting band of our two-band system due to the combined effects of hybridization and  inter-band interactions. We have shown that in the case the former is antisymmetric, the induced superconductivity has a $p$-wave character. This induced pairing corresponds to the $m_{l}=0$ component of the  $l=1$ $p$-wave state and does not break time-reversal invariance.

Our two-band model represents an effective model for a system consisting of a bulk superconductor, described by the $b$-band, and a normal wire deposited on top of it with a non-interacting band of electrons (the $a$-band). The normal wire is coupled to the bulk substrate by an antisymmetric hybridization and an attractive inter-band interaction. We have shown that one component of the induced superconductivity in the wire due to hybridization is of the $p$-wave type, more specifically it is associated with the  time reversal invariant projection, $m_l=0$. We  have studied  the excitations in a wire with this type of induced superconductivity using a Majorana representation. We found that it has a topological superconducting phase that supports  four Majorana modes, two at each end of a finite chain. This is different from Kitaev's spinless chain model that corresponds to the $m_{l}=\pm 1$ $p$-wave state and has a single Majorana at each end of the chain in its non-trivial topological superconducting phase.
Although two Majoranas give rise to a conventional fermion, we have shown using results of previous works that the charge states of the fermion modes at the ends of the $m_l=0$, $p$-wave superconducting chain have highly non-trivial long range correlations. Since these states can be controlled this system may have interesting properties for applications.

\subsection*{Acknowledgments}

We would like to thank Eduardo Miranda, Pedro Sacramento, Claudine Lacroix  and  Tobias Micklitz for useful discussions. This work was supported in part by CNPq - Conselho Nacional de Desenvolvimento Cient\'ifico e Tecnol\'ogico and CAPES - Coordena\c{c}\~{a}o de Aperfei\c{c}oamento de N\'{i}vel Superior (Brazil). M. A. C. also thanks FAPERJ - Funda\c{c}\~ao de Amparo \`a Pesquisa do Estado do Rio
de Janeiro  and H.C to FAPEMIG - Funda\c{c}\~ao de Amparo \`a Pesquisa do Estado de Minas Gerais for partial financial support.


\end{document}